\documentclass[aps,prb,amsmath,amssymb,superscriptaddress,twocolumn,color,epsfig,graphicx,bm,floatfix]{revtex4-1}
\usepackage[super]{nth}
\usepackage{amsmath}    
\usepackage{mathtools}
\usepackage{siunitx}
\usepackage{multirow}
\usepackage{array}
\usepackage{tabulary}
\usepackage{longtable}
\usepackage[version=4]{mhchem}
\newcolumntype{K}[1]{>{\centering\arraybackslash}p{#1}}

\usepackage{graphicx}  
\usepackage{verbatim}  
\usepackage{color}     
\usepackage{subfigure} 
\usepackage{hyperref}  
\begin{document}

\title{Thermal conductivity of \ce{CaF2} at high pressure} 
\author{Somayeh Faraji}
\affiliation{Department of Physics, Institute for Advanced Studies in Basic Sciences, P.O. Box 45195-1159, Zanjan, Iran}
\affiliation{Department of Physics, University of Tehran, P.O. Box: 14395/547, Tehran, Iran}
\author{S. Mehdi Vaez Allaei}
\affiliation{Department of Physics, University of Tehran, P.O. Box: 14395/547, Tehran, Iran}
\author{Maximilian Amsler}
\email{amsler.max@gmail.com}
\affiliation
{Department of Chemistry and Biochemistry, University of Bern, Freiestrasse 3, CH-3012 Bern, Switzerland}

\date{\today}
\begin{abstract}
We study the thermal transport properties of three
\ce{CaF2} polymorphs up to a pressure of 30~GPa using first-principle calculations 
and an interatomic potential based on machine learning.
The lattice thermal conductivity $\kappa$ is computed by 
iteratively solving the linearized Boltzmann 
transport equation (BTE) and by taking into account three-phonon scattering. 
Overall,  $\kappa$ increases nearly linearly with pressure,
and we show that the recently discovered $\delta$-phase with $P\bar{6}2m$ symmetry and the previously 
known $\gamma$-\ce{CaF2} high-pressure phase have significantly 
lower lattice thermal conductivities than the ambient-thermodynamic cubic fluorite ($Fm\bar{3}m$) structure.
We argue that the lower $\kappa$ of these two high-pressure phases
stems mainly due to a lower contribution of acoustic modes to $\kappa$
as a result of their small group velocities.
We further show that the phonon mean free paths are very short for the $P\bar{6}2m$ and $Pnma$ structures at high temperatures, and
resort to the Cahill-Pohl model to assess the lower limit of thermal conductivity in these domains.

\end{abstract}

\maketitle
\section{Introduction\label{sec:introduction}}
Calcium fluoride (\ce{CaF2}) has a variety of technological applications
due to its remarkable optical properties and  
its high thermal stability~\cite{liang2015,cazorla2014,lyberis2012,sang2011}. 
At ambient conditions, $\alpha$-\ce{CaF2} crystallizes in the 
cubic fluorite structure with $Fm\bar{3}m$ symmetry.
In this structure, \ce{CaF2} exhibits a superwide band gap of $12$~eV with
excellent light transmission over a wide spectrum, and 
a high laser damage threshold. 
These properties render \ce{CaF2} an ideal candidate for optical windows,
main lens substrates in large scale semiconductor 
micro-lithography systems, and photo-detectors~\cite{liberman1999materials,daimon2002high,wagner2010euv,liang2015ultralow,luo2012single,sang2011high}.

Cubic \ce{CaF2} undergoes a sequence of structural 
phase transitions at increased pressures~\cite{Seifert1966,P1,Gerward,Dorfman,Speziale,P3,Wu,Shi2009,Elkin}. Above 8--10~GPa,  \ce{CaF2} transforms to the denser orthorhombic cotunnite $\gamma$-phase
with  $Pnma$ symmetry, 
accompanied by an increased coordination number of Ca from $8$ to $9$.  
X-ray diffraction and Raman spectroscopy have shown that this high-pressure phase 
is stable up to $49$~GPa at room temperature~\cite{Gerward,Speziale}. 
As pressure increases further, the stability of $\gamma$-\ce{CaF2} decreases, 
and above 72~GPa, a further transition occurs to a hexagonal $P6_{3}/mmc$ phase~\cite{dorfman2010phase}.

In addition to these experimentally observed low-temperature high-pressure phases,
high-temperature modifications thereof have been studied predominantly using 
computational models. Using \textit{ab initio} structural searches,
Nelson~\textit{et al.} recently proposed a hypothetical structure with $P\bar{6}2m$ symmetry
as a high-temperature polymorph of $\gamma$-\ce{CaF2},
referred to as  $\delta$-\ce{CaF2}~\cite{nelson2017high}.
Similar to its ambient-pressure counterpart,
$\delta$-\ce{CaF2} is predicted to undergo a transition to a 
superionic phase with bcc structure at temperatures exceeding $\approx2500$~K
at 20~GPa.

Despite these theoretical studies, little is known about the high-pressure behavior of the thermal transport properties in \ce{CaF2} polymorphs. At ambient pressure, the lattice thermal conductivity of $\alpha$-\ce{CaF2} has been 
studied both through experiments and
computations. 
In two separate early experiments in 1957 an 1960, the near room-temperature value of the lattice thermal conductivity was measured 
to be $5.5$~\cite{charvat1957thermal} and $9.5$~\cite{mccarthy1960thermal} ~$\text{Wm}^{-1}\text{K}^{-1}$, respectively. 
Later, Slack~\cite{slack1961thermal}
reported a value of $11.69$~$\text{Wm}^{-1}\text{K}^{-1}$ in 1961.
Theoretical room-temperature values from simulations have been predicted in the range of 7.0 and 8.6~$\text{Wm}^{-1}\text{K}^{-1}$.~\cite{lindan1991molecular,qi2016lattice}.
To the best of our knowledge, the only work on the pressure dependence of $\kappa$
in $\alpha$-\ce{CaF2} 
was reported $9.7<\kappa<10.9$~$\text{Wm}^{-1}\text{K}^{-1}$,
measured using a dynamic two-strip method at room temperature in a narrow pressure range of $0.1$ to $1.0$ GPa.~\cite{Andersson_1987}.

In this work, we study the thermal conductivity of the $\alpha$, $\gamma$, and $\delta$ phases of \ce{CaF2}
as a function of pressure in the range of $0$-$30$~GPa.
To alleviate the computational burden of \textit{ab initio} calculations, we resort to training an efficient machine learning interatomic potential to accelerate the assessment of the lattice thermal conductivity $\kappa$. 
We show that the value of $\kappa$ for the 
$\gamma$-\ce{CaF2} and $\delta$-\ce{CaF2} phases are 
lower than that of the $\alpha$ phase across the whole pressure domain. 
In particular, the extremely small phonon mean-free-paths
in these two phases leads to a potential break-down of the 
Boltzmann transport equation (BTE). Hence, we assess the validity of the BTE results based on the amorphous limit using the Cahill-Pohl model and draw the associated temperature-pressure transition boundary.

\section{Methods}

\subsection{Interatomic Potential} 
We use CENT, a neural-network-based interatomic potential, that takes into account charge transfers to model the ionic bonding in \ce{CaF2}.~\cite{Ghasemi_ANN} 
The construction of the \ce{CaF2} CENT potential is discussed in detail elsewhere~\cite{faraji2017high}, and we employ its implementation in the
\textit{FLAME} package~\cite{amsler2020flame,Asna2017,Rasoulkhani2017}. The 
particular parametrization of our CENT potential has been used elsewhere to obtain physical properties of \ce{CaF2}~\cite{faraji2017high}
and to study surface morphologies of \ce{CaF2}~\cite{faraji2019surface}. 

\subsection{Density Functional Theory}
Structural relaxations and single-point total energy calculations were performed with density functional theory (DFT) calculations  
at selected pressures, 2, 10, and 30~GPa. 
We used the plane-wave Quantum ESPRESSO simulation package~\cite{QE-2009,QE-2017} in conjunction with the Perdew-Burke-Ernzerhof (PBE)~\cite{perdew1996} parametrization of the exchange-correlation functional and ultrasoft pseudopotentials.~\cite{PhysRevB.41.7892} 
The wave functions and electron densities were expanded with a plane wave basis set up to a kinetic cutoff energy of $45$ Ry and $540$ Ry, respectively. 
The Brillouin zone was sampled using $16\times16\times16$, $14\times14\times16$, $14\times16\times14$ Monkhorst-Pack~\cite{PhysRevB.13.5188} k-points meshes for $\alpha$-\ce{CaF2}, $\delta$-\ce{CaF2}, and $\gamma$-\ce{CaF2}, respectively. 
The atomic positions were relaxed until the maximal force acting on the atoms was less than $1\times10^{-5}$ Ry/Bohr.

\subsection{Phonons\label{sec:lattice_dynamics}} 
The second order interatomic force constants were calculated by the finite difference approach
using the supercell method as implemented
in the Phonopy package.~\cite{togo2015first}.
Supercells of dimension $4\times4\times4$, $2\times3\times2$,
and $2\times2\times4$ were used for $\alpha$-\ce{CaF2}, $\gamma$-\ce{CaF2},
and $\delta$-\ce{CaF2}, respectively,
leading to cells consisting of $192$, $144$, and $144$ atoms. 
A finite difference step size of $0.01$~\AA\,was applied to displace the atoms.
A $\textbf{q}$-point mesh of $50\times50\times50$ was used for the BZ integration.

\subsection{BTE thermal transport\label{subsec:thermal_transport}}
The thermal transport calculations were carried out by taking into account 
anharmonic three-phonon interactions. 
Third-order force constants were computed from finite differences using
the supercell method with the same sizes
used in the calculations of the second-order force constants.
Atomic displacements were created
using the \texttt{thirdorder.py} script
included in the ShengBTE distribution,~\cite{li2014shengbte} taking into account up to the \nth{5}-nearest neighbors
to truncate the three-body interactions, which gives well converged values of $\kappa$.
The thermal conductivity in the BTE is given by ~\cite{luo2013gallium} 
\begin{equation}
\mathbf{\kappa} = \frac{1}{3VN_{q}}\sum_{q\gamma}
C_{q\gamma}\mathbf{v}_{q\gamma}^{2} \tau_{q\gamma}
    \label{eqn:kappa_alphabeta}
\end{equation}
where $V$ is the volume of the cell containing N atoms, 
$q$ refers to the wave vector in the first Brillouin zone, 
$N_{q}$ is the number of discrete q-points, 
$\gamma$ is the mode index that refers to different phonon branches, 
$C_{q\gamma}$ denotes the mode specific heat capacity at constant volume, 
$\mathbf{v}$ is the phonon group velocity ($\mathbf{v}_{q\gamma}=\nabla_{q}\omega_{q\gamma}$), 
$\tau$ is the phonon life time, which is related to the 
MFP $\mathbf{\lambda}=\mathbf{v}\cdot{\tau}$. 

The detailed effects of the cutoff-distance on the thermal conductivities
are shown in section S3
of the supporting information.
The second- and third-order interatomic force constants
were fed into the ShengBTE package 
to calculate $\kappa$ by iteratively solving the linearized phonon Boltzmann transport equation 
for temperatures ranging from $100$~K to $900$~K. 
Both isotopic and three-phonon scattering were considered. 
The isotopic scattering rates were calculated by applying 
the Pearson deviation coefficients 
incorporated in ShengBTE. 
The so-called proportionality constant \textit{scalebroad},   
related to the adaptive Gaussian broadening technique,  
was set to $0.2$ in all the ShengBTE calculations,
together with
with 19$\times$19$\times$19, 19$\times$19$\times$19, and 15$\times$15$\times$12  $\mathbf{q}$-point grids
for $\alpha$-\ce{CaF2}, $\delta$-\ce{CaF2}, and $\gamma$-\ce{CaF2}, respectively.

\subsection{Cahill-Pohl model}
We estimate the amorphous limit of the thermal conductivity $\kappa_\text{CP}$
using the Cahill-Pohl model~\cite{cahill1988lattice}, which
is an extension of the Einstein model.  
While Einstein assumed that the thermal energy is transported 
between neighboring atoms vibrating with a single frequency, 
the Cahill-Pohl model proposes that the energy is transferred 
between collective vibrations. 
Therefore, the model includes a range of frequencies, 
instead of a single frequency used by Einstein. 
In this model, the thermal conductivity is expressed as follows
(details in Ref.[\onlinecite{kaviany2014heat}])
\begin{eqnarray}
    \kappa_\text{CP}&=&(\frac{\pi}{6})^{\frac{1}{3}}k_{B}n^{\frac{2}{3}}
    \sum_{\alpha}v_{\alpha}(\frac{T}{T_{\alpha}^{D}})^{2}\int_{0}^{T_{\alpha}^{D}/T}\frac{x^{3}e^{x}}{(e^{x}-1)^{2}}dx,\nonumber\\
    x&=&\frac{T_{\alpha}^{D}}{T},
    \label{eq:Cahill-Pohl}
\end{eqnarray}
where $n$ is the density of the atoms in the solid ($m^{-3}$), 
$v_{\alpha}$ is the low-frequency speed of sound (from acoustic phonons) 
for polarization $\alpha$.
$T_{\alpha}^{D}=\frac{\hbar}{k_{B}}v_{\alpha}(6n\pi^{2})^{\frac{1}{3}}$ is the characteristic temperature equivalent to the Debye temperature for that polarization which corresponds to the activation of all phonons.  
$x$ is the reduced phonon energy and the summation runs over the vibrational polarizations
(one longitudinal and two transverse acoustic branches). 
The $v_{\alpha}$ of each acoustic group velocity was determined using harmonic lattice dynamics from Sec.~\ref{sec:lattice_dynamics}. 

\section{Results and discussion} 

We start out by validating the quality and predictive power of our CENT potential
with respect to DFT results based on the three relevant phases $\alpha$-\ce{CaF2},  $\gamma$-\ce{CaF2}, and $\delta$-\ce{CaF2}.
The thermodynamic properties including the transition pressures
are well reproduced by CENT. As shown in Fig.~S1 of the supplementary materials,
the phase transition from $\alpha$-\ce{CaF2} to $\gamma$-\ce{CaF2}
occurs at $8$~GPa which is close to experimental measurements. 
Also, the enthalpy differences of the $\delta$-\ce{CaF2} and 
$\gamma$-\ce{CaF2} decreases with increasing pressure. In fact, these two high pressure phases are energetically very close to each other, i.e., dropping from 12.7 to 2.1~meV/atom in the pressure range between 2 and 30~GPa. 
We then compare the dynamical properties predicted by the CENT potential
with DFT values at 2~GPa. The phonons arising from the CENT potential as well as
the phonon DOS agree well with the results from DFT (see Fig.~S2 in the supplementary materials). Similarly, the lattice thermal
conductivities from CENT are in excellent agreement with the DFT predictions,
as shown in Fig.~S3 in the supplementary material.

\begin{figure}[!htbp]
\centering 
\includegraphics[width=0.48\textwidth]{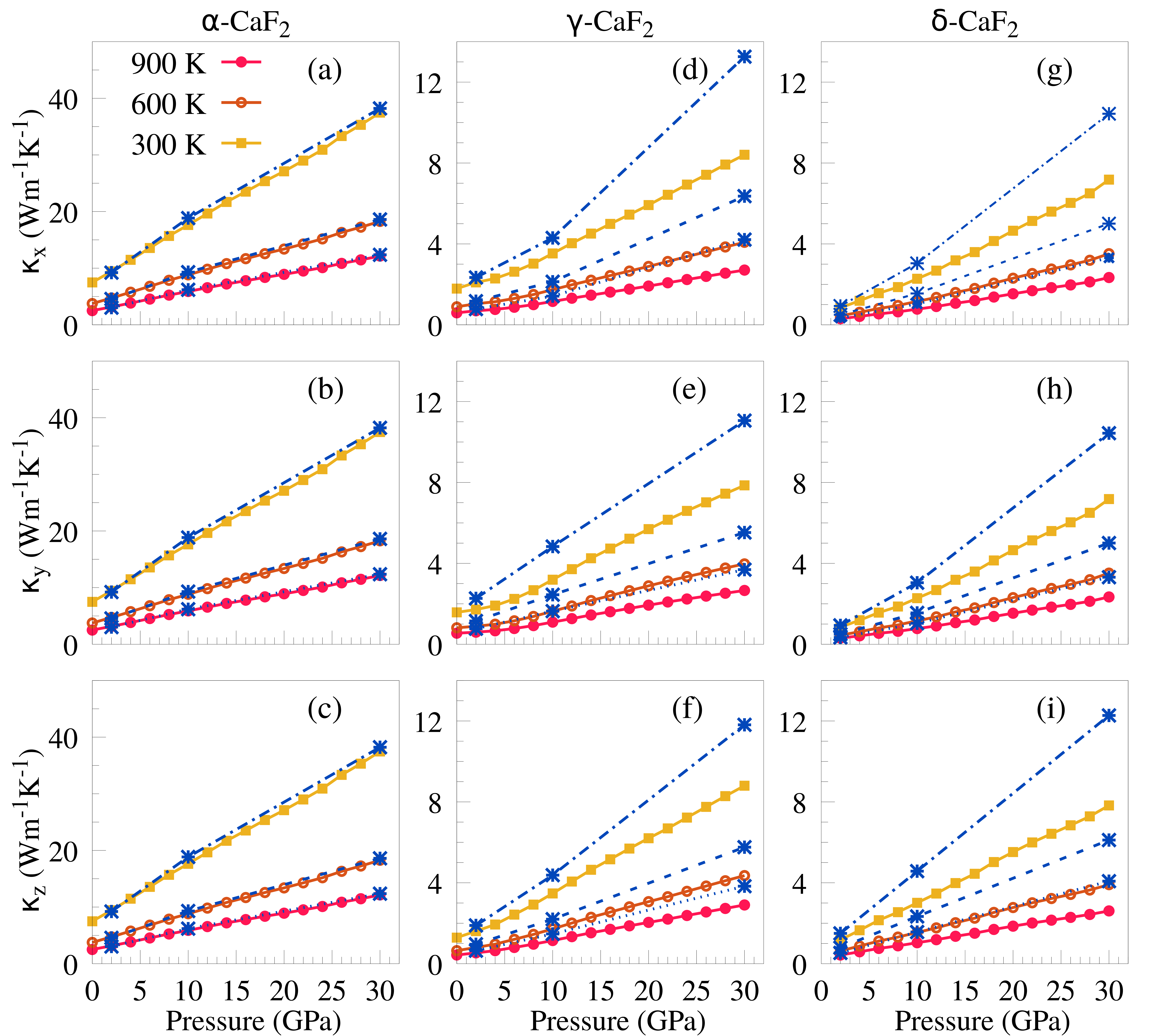}
    \caption{(color online)
    The components of the
    lattice thermal conductivities  $\kappa_{x,y,z}$ 
    of $\alpha$-\ce{CaF2} (panels (a) to (c)), 
    $\gamma$-\ce{CaF2} (panels (d) to (f)), 
    and $\delta$-\ce{CaF2} (panels (g) to (i)), 
    as a function of pressure at temperatures of $300$, $600$, and $900$~K. 
    DFT data are shown with blue crosses, while 
    CENT results are shown with yellow squares ($300$~K), orange circles ($600$~K), 
    and pink dots ($900$~K). 
    } 
\label{fig:kappa_pressure}
\end{figure}

Next, we study the evolution of the thermal conductivity of the
three phases as a function of pressure.
Since the $\delta$-\ce{CaF2} phase exhibits imaginary
phonon modes at 0~GPa, we focus on the pressure regime between
2 and 30~GPa within which all structures
are dynamically stable.
Fig.~\ref{fig:kappa_pressure}
plots the components $\kappa_{x,y,z}$ of the thermal conductivity at selected temperatures, and
shows that their values increase almost linearly with pressure. 
This increase in thermal conductivtiy  can be
rationalized in a first approximation by the
decrease in volume $V$ in the denominator in
Eq.~\eqref{eqn:kappa_alphabeta}
as the pressure increases.
The room-temperature thermal conductivity
of all three phases at different pressures and room temperature are also summarized in Table~\ref{tab:kappa} 

\begin{table*}[!htbp]
\small
\caption{
    The components of the lattice thermal conductivity $\kappa_{x,y,z}$ using CENT in units of ~$\text{Wm}^{-1}\text{K}^{-1}$
    at $300$~K and at selected pressures,
    together with available values from the literature. 
    Results from DFT calculations are given in paranthesis. 
\label{tab:kappa} }
\begin{tabular*}{1.0\textwidth}{@{\extracolsep{\fill}}lllllll}
\hline
    Phase          &    Components of $\kappa$                                       & 0~GPa & 2~GPa         &10~GPa          &20~GPa  &30~GPa \\\hline\hline
    $\alpha$-\ce{CaF2}   & $\kappa_{x}=\kappa_{y}=\kappa_{z}$  &7.5    & 9.4   (9.2) &17.6   (18.8) &27.1    & 37.5   (38.2) \\
                   &                &  7.04~\cite{plata2017efficient}, 8.6~\cite{phonon_qi2016lattice} , 7.0 $\pm{0.39}$~\cite{lindan1991molecular} &   &   &  & \\
                   &                                           &       &               &                &        &        \\
    $\delta$-\ce{CaF2}   & $\kappa_{x}=\kappa_{y}$               &       & 0.9   (0.9) & 2.3   (2.9)  & 4.6    & 7.2   (10.0 ) \\
                   & $\kappa_{z}$                            &       & 1.2   (1.5) &3.0   (4.5)   & 5.5    & 7.8   (12.4)\\
                   &                                           &       &               &                &        &               \\
                   & $\kappa_{x}$                            & 1.8   & 2.1   (2.3) & 3.5   (4.1)  & 5.9    & 8.4   (12.3) \\
    $\gamma$-\ce{CaF2}         & $\kappa_{y}$                            & 1.6   & 1.7   (2.3) & 3.2   (4.8)  & 5.7    & 7.9   (10.8) \\
                   & $\kappa_{z}$                            & 1.3   & 1.6   (1.9) & 3.5   (4.6)  & 6.2    & 8.8   (12.5) \\\hline
\end{tabular*}
\end{table*}
At all pressures, $\alpha$-\ce{CaF2}  
has a significantly higher value of $\kappa$ than any of the other two phases at a given temperature. 
Table~\ref{tab:kappa} also contains  
the room-temperature zero-pressure value of the 
thermal conductivity, $\kappa_\text{ambient}$,
of  $\alpha$-\ce{CaF2} from other theoretical studies in the literature. 
We obtain $\kappa_\text{ambient}=7.5$ ~$\text{Wm}^{-1}\text{K}^{-1}$, 
which is close to the value of $7.0$ ~$\text{Wm}^{-1}\text{K}^{-1}$ reported 
by Plata \textit{et al}.~\cite{plata2017efficient} 
In comparison with experimental results from Andersson ~\textit{et al}, 
our value of $\kappa_\text{ambient}$ is about $2.2$ ~$\text{Wm}^{-1}\text{K}^{-1}$ 
lower than the experimental measurement of $9.7$ ~$\text{Wm}^{-1}\text{K}^{-1}$ through 
a two-strip method.~\cite{Andersson_1987} 

Our values of $\kappa$ for $\delta$-\ce{CaF2} and $\gamma$-\ce{CaF2}  
show that, unlike $\alpha$-\ce{CaF2}, these two phases exhibit slight anisotropies along their three components.
At $300$~K, the components of $\kappa_{x,y,z}$
for the $\gamma$-\ce{CaF2} and $\delta$-\ce{CaF2} phases at $2$~GPa are approximately 
$\{2.1,1.7,1.6$\} and $\{0.8, 0.8, 1.2$\}~$\text{Wm}^{-1}\text{K}^{-1}$, respectively, while $\kappa$ itself are 1.8 and $0.93~\text{Wm}^{-1}\text{K}^{-1}$ for $\gamma$-\ce{CaF2} and $\delta$-\ce{CaF2}, respectively.  
The very low thermal conductivity of the $\delta$-phase at low pressures can be primarily attributed to the 
soft acoustic phonon mode along K--$\Gamma$
in the first Brillouin zone.
(see section S4 and Fig.~S5 for in the SI).

There are several factors leading to the deceased $\kappa$  of 
$\gamma$-\ce{CaF2} and $\delta$-\ce{CaF2} compared to the cubic structure.
Eq.~\eqref{eqn:kappa_alphabeta}
contains the product of 
heat capacity, 
phonon group velocity, and phonon mean free path,
the effects of which we can study individually.
We first investigate the heat capacities per unit volume 
at selected pressure and temperatures, and show its evolution in Fig.~S4 of the SI.
The heat capacity rapidly increases with 
temperature $T$, and is proportional to T$^{3}$ at low $T$, 
whereas it tends to a constant value at a high temperature, 
following the Dulong-Petit law. 
In the case of $\alpha$-\ce{CaF2}, the obtained value for C$_{v}$ 
at zero pressure and at temperature $300$~K is $65.57$ J/m/K, which is comparable  
with the experimental value of 67.11 J/m/K.~\cite{Andersson_1987}  
The obtained values of C$_{v}$ at temperature $300$~K and pressure $2$~GPa 
for $\alpha$-\ce{CaF2}, $\delta$-\ce{CaF2}, and $\gamma$-\ce{CaF2} 
are $64.92$, $65.10$, and $64.64$ J/m/K, respectively.  
The value of $C_{v}$ decreases with increasing pressure  
(see insets in Fig.~S4 of the SI) at given temperature, and at $30$~GPa reaches $58.29$, $59.77$, and $59.54$ J/m/K for $\alpha$-\ce{CaF2}, $\delta$-\ce{CaF2}, and $\gamma$-\ce{CaF2}, respectively.
Overall, the difference in C$_{v}$ among the three phases
is minute (within less than 3~\%) and cannot account for
the strong deviations of $\kappa$.

\begin{figure}[!htbp]
\centering
\includegraphics[width=0.48\textwidth]{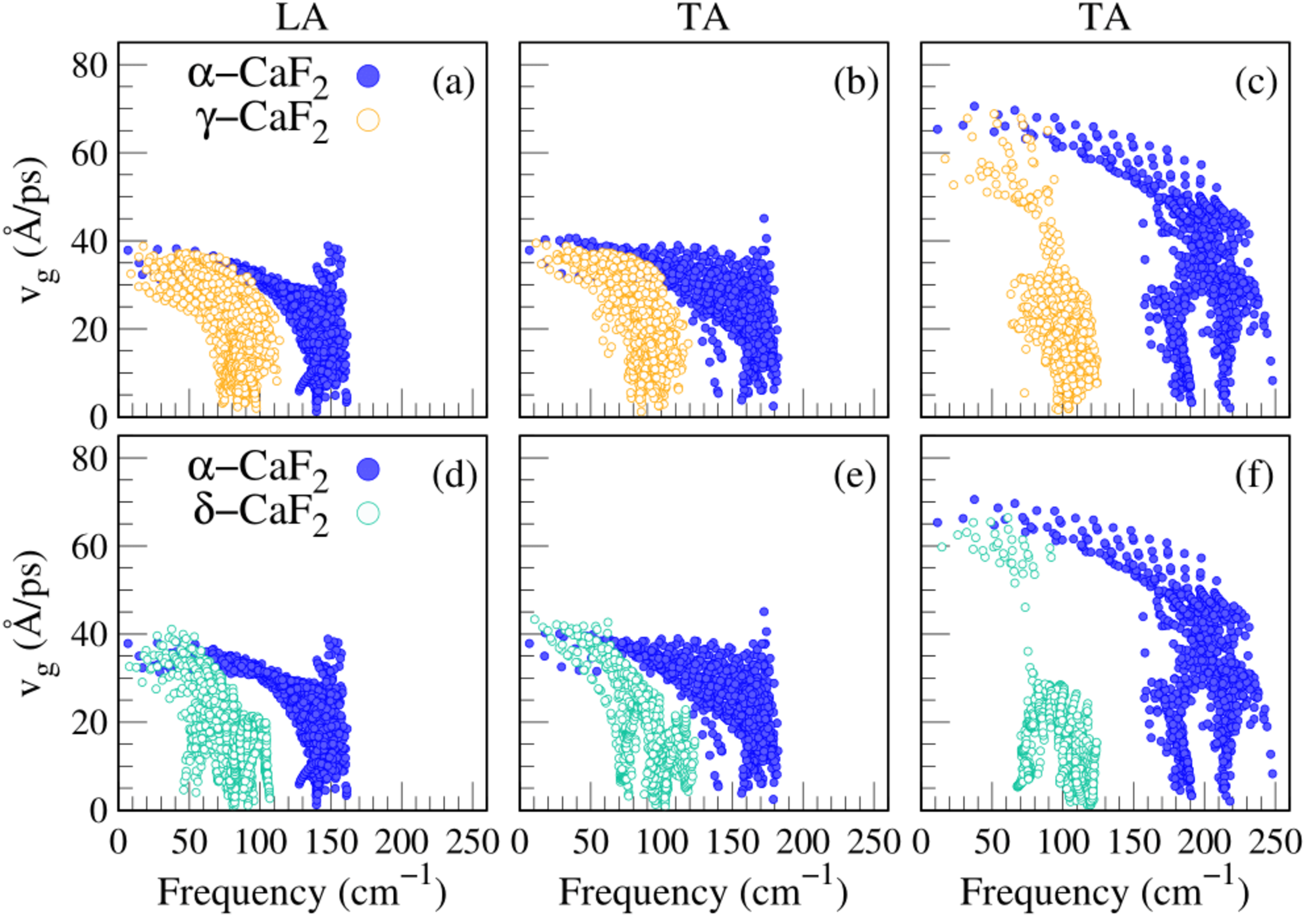}
    \caption{(color online) Group velocities of the longitudinal and transversal acoustic modes 
    as functions of frequency for the
    $\gamma$-\ce{CaF2} ((a) through (c)) and $\delta$-\ce{CaF2} ((d) through (f))  
    in comparison with $\alpha$-\ce{CaF2}   
    at $300$~K and $2$~GPa.} 
\label{fig:group_acoustic_p2GP}
\end{figure}

We now turn our attention to the group velocities $v_g$ of the acoustic phonon modes, which are in general responsible for  a large fraction of the thermal transport. 
Fig.~\ref{fig:group_acoustic_p2GP} shows $v_{g}$ of the longitudinal and transverse acoustic (LA and TA) branches 
of $\gamma$-\ce{CaF2} and $\delta$-\ce{CaF2}, plotted on top of the values of $\alpha$-\ce{CaF2}.
Note that the $\delta$-phase exhibits a particularly 
soft acoustic branch with a low $v_{g}$ along K--$\Gamma$ (see Fig.~S5 in the SI). 
Overall, $\alpha$-\ce{CaF2} has 
larger group velocities than either $\gamma$-\ce{CaF2} or $\delta$-\ce{CaF2}.
To quantify the difference in the group velocities, we consider the mean values of the LA and the two TA modes,
$\bar{v}_g^{\text{LA}}$, $\bar{v}_g^{\text{TA}_1}$, and $\bar{v}_g^{\text{TA}_2}$.
The ratios of these average velocities of $\alpha$-\ce{CaF2}
with respect to the $\gamma$ and $\delta$-phases
is
$\{\bar{v}_g^{\text{LA}}, \bar{v}_g^{\text{TA}_1}, \bar{v}_g^{\text{TA}_2}\}_{\alpha}/\{\bar{v}_g^{\text{LA}}, \bar{v}_g^{\text{TA}_1}, \bar{v}_g^{\text{TA}_2}\}_{\gamma}=\{1.7, 1.7, 1.9\}$, 
and 
$\{\bar{v}_g^{\text{LA}}, \bar{v}_g^{\text{TA}_1}, \bar{v}_g^{\text{TA}_2}\}_{\alpha}/\{\bar{v}_g^{\text{LA}}, \bar{v}_g^{\text{TA}_1}, \bar{v}_g^{\text{TA}_2}\}_{\delta}=\{1.7, 1.6, 1.93\}$. Hence, the group velocities of $\alpha$-\ce{CaF2} is
almost twice as high as the corresponding values
in $\gamma$-\ce{CaF2} and $\delta$-\ce{CaF2}.

\begin{figure}
    \centering
    \includegraphics[width=0.49\textwidth]{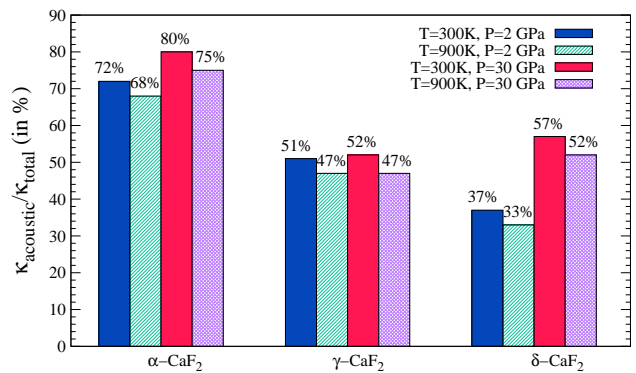}
    \caption{The fraction of of acoustic modes $\kappa_\text{acoustic}$ contributing to the total thermal conductivity  $\kappa_\text{total}$
    at pressures of $2$ and $30$~GPa and temperatures of $300$ and $900$~K for the three relevant \ce{CaF2} phases.}
    \label{fig:band_contribution}
\end{figure}

Further, the contributions of acoustic modes to the thermal transport
is influenced by their interaction with the 
optical modes, i.e., the amount of
heat that is scattered through optical phonons.
In general, phases with larger, complex structures 
tend to have larger contributions from optical scattering,
with stronger coupling between acoustic and optical
modes.
Fig.~\ref{fig:band_contribution} shows the fraction of
acoustic modes contributing to the
total thermal conductivity, $r_\kappa=\kappa_\text{acoustic}/\kappa_\text{total}$ for the $\alpha$,  $\gamma$, and $\delta$ phase.
At any pressure and temperature,
$\alpha$-\ce{CaF2} exhibits the largest value of
$r_\kappa$.
Both $\gamma$-\ce{CaF2} and $\delta$-\ce{CaF2}
show strong contributions of optical phonon scattering,
in particular for $\delta$-\ce{CaF2}
at low lower pressures.
Again, this behavior can be attributed 
to the soft-mode in one of the acoustic branches 
of $\delta$-\ce{CaF2} at 2~GPa, which becomes
less pronounced with increasing pressure as shown
in Fig.~S5 in the SI.

\begin{figure}[!htbp]
\centering
\includegraphics[width=0.48\textwidth]{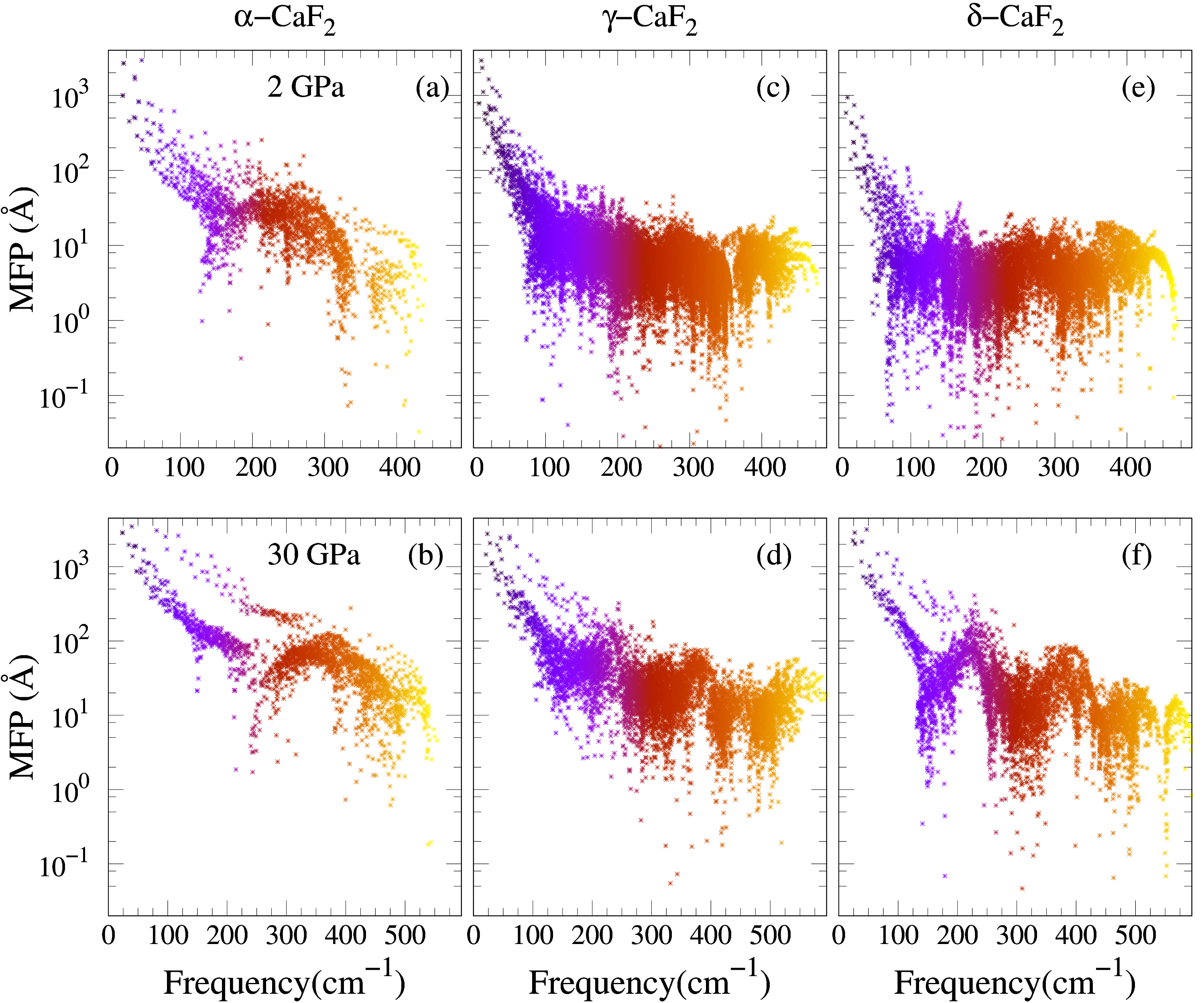}
    \caption{(color online) Phonon mean free path (MFP) 
    of all phonon modes as a function of frequency at pressures of $2$~GPa (first row) 
    and $30$~GPa (second row)  
    of $\alpha$-\ce{CaF2} (panels (a) and (b)), $\gamma$-\ce{CaF2} (panels (c) and (d)), 
    and  $\delta$-\ce{CaF2} (panels (e) and (f)).
    }
\label{fig:MFP_Freq}
\end{figure}

We also compare the phonon MFP in
Fig.~\ref{fig:MFP_Freq} at 2 and 30~GPa at a temperature of $300$~K. 
Overall, the MFPs of the $\alpha$-phase
are longer than either of the high-pressure phases.
In fact, the MFP of a significant fraction of modes are
shorter than the average inter-atomic distance of $\approx 2.4~$\AA\, in both the $\gamma$ and $\delta$-phases,
leading to an inaccurate description
of thermal transport within the BTE
by dramatically underestimating the value of $\kappa$~\cite{allen1993thermal}.

\begin{figure}[!htbp]
\centering
\includegraphics[width=0.48\textwidth]{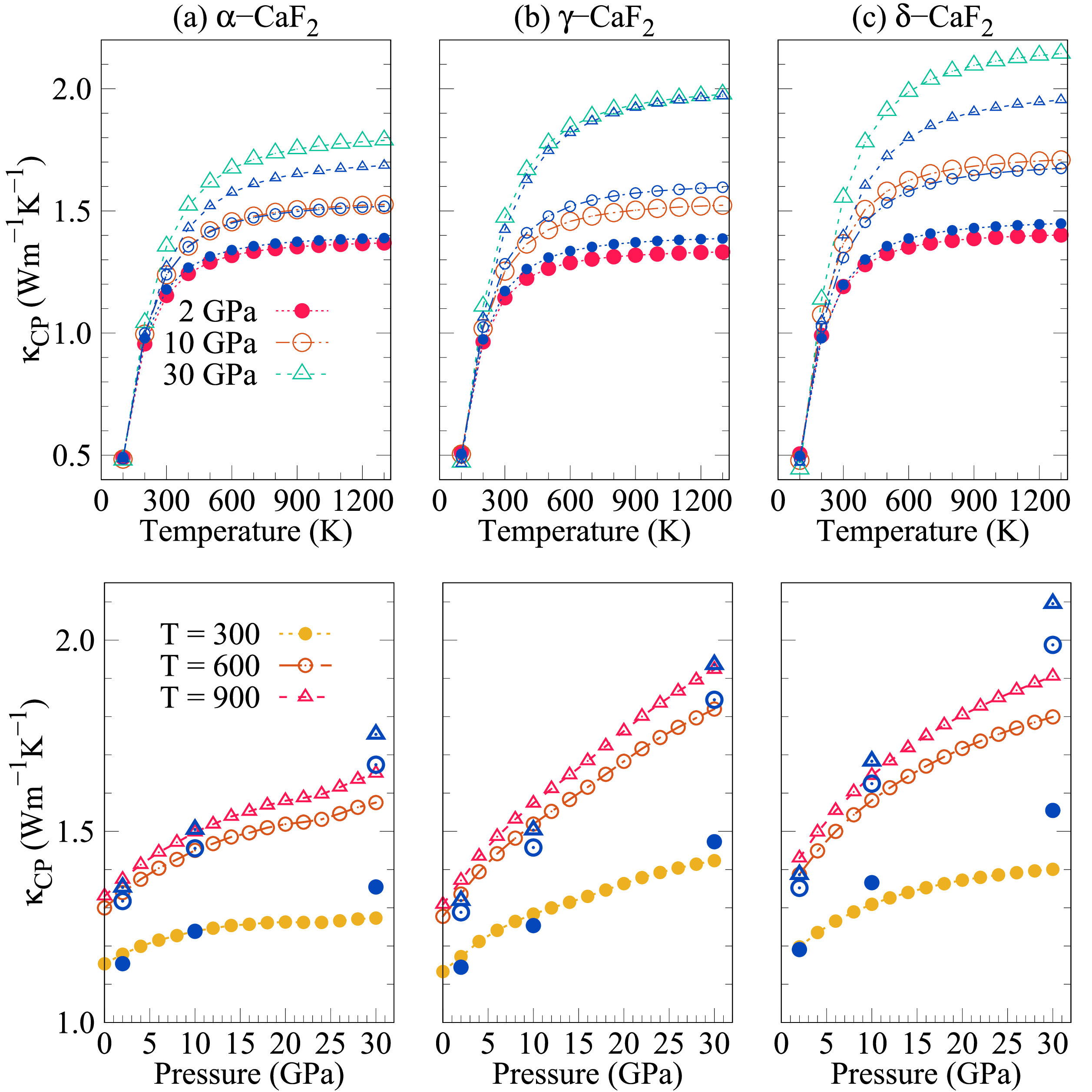}
    \caption{(color online) The amorphous limit of the thermal conductivity $\kappa_\text{CP}$ based on the Cahill-Pohl model of
    (a) $\alpha$-\ce{CaF2}, (b) $\gamma$-\ce{CaF2}, and (c) $\delta$-\ce{CaF2} based on CENT calculations
    as a function of temperature at selected pressures (first row) 
    and as a function of pressure for temperatures $300$, $600$, and $900$~K (second row). 
    DFT  results at $2$, $10$, $30$~GPa are shown
    with blue symbols.
      }
\label{fig:kappa_CP}
\end{figure}

To address this issue, we assess the limitations of the 
BTE by comparing its results to the
Cahill-Pohl model, which provides an 
estimate of the lower bound in the amorphous limit, $\kappa_\text{CP}$.
Fig.~\ref{fig:kappa_CP} shows the values of $\kappa_\text{CP}$ 
as a function of temperature and pressures for 
the $\alpha$, $\gamma$, and $\delta$-\ce{CaF2}.
We observe two very clear trends: 
(a) $\kappa_\text{CP}$ increases with temperature
at a given pressure, plateauing out above 
$\approx 600$~K (see top row in Fig.~\ref{fig:kappa_CP}),
and (b) $\kappa_\text{CP}$ increases steadily with pressure
at constant temperature (see bottom row in Fig.~\ref{fig:kappa_CP}). The values of $\kappa_\text{CP}$
are particularly high for $\delta$-\ce{CaF2}, 
which indicates that an especially large error can be
expected in the BTE model.

\begin{figure}[!htbp]
    \centering
    \includegraphics[width=0.48\textwidth]{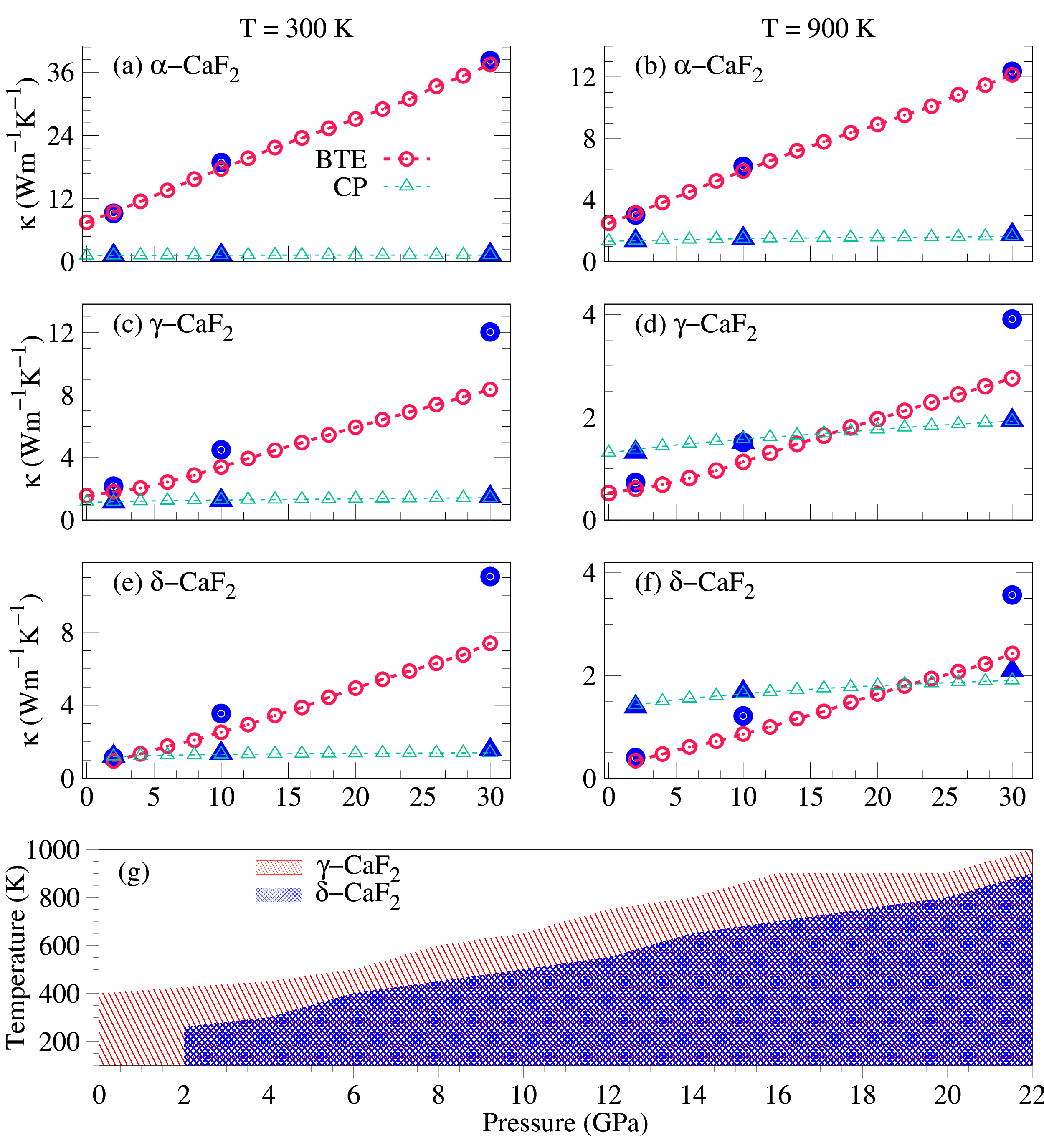}
    \caption{(color online) 
    The thermal conductivity $\kappa$ 
    as a function of pressure 
    for  $\alpha$-\ce{CaF2} ((a) and (b) ), 
    $\gamma$-\ce{CaF2} ((c) and (d)), 
    and $\delta$-\ce{CaF2} ((e) and (f)) 
    Cahill-Pohl model (green triangles) at $300$ and $900$~K 
    together with the results of BTE (red circles) at the same conditions.  
    The values obtained from DFT results are shown with blue symbols. 
    The calculated transition boundary where the 
    $\kappa_\text{CP}$>$\kappa_\text{BTE}$
    for $\gamma$-\ce{CaF2} 
    and $\delta$-\ce{CaF2} are shown in panel (g), 
    where the shaded regions indicate the $T-p$-range where
    BTE is reliable. 
    }
    \label{fig:kappa_CP_BTE}
\end{figure}

To assess the limits of the BTE, we map out
the boundary in $T$ and $p$ where $\kappa_\text{BTE}$
drops below the amorphous limit, $\kappa_\text{CP}$.
Fig~\ref{fig:kappa_CP_BTE} plots the $\kappa_\text{BTE}$ and
$\kappa_\text{CP}$ at selected temperatures as a function
of pressure. 
For $\alpha$-\ce{CaF2}, the thermal conductivities predicted through the BTE are reliable, as their values
remain above $\kappa_\text{CP}$ for all pressures and temperatures
considered here. However, the BTE breaks down for $\delta$-\ce{CaF2} and $\gamma$-\ce{CaF2}, especially
at low temperatures and low pressures.
The transition boundary where $\kappa_\text{BTE}$ crosses 
$\kappa_\text{CP}$ in $T$ and $p$ is mapped out in Fig.~\ref{fig:kappa_CP_BTE}(g), 
showing that BTE only yields
reliable results within the regime of high pressure and low temperatures.


\section{Conclusions}
In summary, we studied the effect of pressure and temperatures on the thermal transport properties
of three crystalline \ce{CaF2} phases, using DFT and a machine-learning based interatomic potential.
Our results show that the two high-pressure phases,  
$\delta$-\ce{CaF2} and $\gamma$-\ce{CaF2}, exhibit significantly 
lower thermal conductivities $\kappa$
than the cubic $\alpha$-phase. 
We argue that the source of this large difference in
$\kappa$ stems from lower group velocities
of the acoustic modes, and 
the larger contributions of phonon scattering events 
involving the optical modes in the $\delta$ and $\gamma$-phase which additionally impedes the transport of heat.
A careful analysis of the phonon 
scattering shows that
the MFPs (and the associated phonon lifetimes) 
are extremely short for the $\delta$ and $\gamma$-phases,
leading to the low values of $\kappa$.
In fact, for high temperatures and at low pressures the
MFPs are so short that they drop below the mean atomic bond lengths,
and we expect that the thermal conductivity 
will eventually converge to the amorphous limit which
we estimate using the Cahill-Pohl model.
Despite these limitations, our results show that
the high-pressure phases exhibit 
around a factor of 5 times lower thermal conductivity
than the ambient ground state.

\section{Acknowledgments}
MA acknowledges support from the Swiss National Science Foundation (project P4P4P2-180669).


\bibliography{ann_CaF2_thermal}
\end{document}